\documentclass[letterpaper,titlepage,10pt]{article}

\usepackage{hyperref}

\usepackage{amssymb,amsmath,amsfonts}
\usepackage{epsfig}

\def\clock{{\count0=\time
           \divide\count0 60
           \ifnum\count0<10 0\fi\the\count0
           \multiply\count0 -60 \advance\count0 \time
           :\ifnum\count0<10 0\fi \the\count0
         }}
\newcommand{\timestamp}{{\small\vbox{\hbox{\tt\jobname.tex}
\hbox{\the\day/\the\month/\the\year, \clock}}}}

\newcommand{\CL}{\mathcal{L}}
\newcommand{\CO}{\mathcal{O}}
\newcommand{\CN}{\mathcal{N}}

\newcommand{\Z}{\mathbb{Z}}
\newcommand{\C}{\mathbb{C}}
\newcommand{\R}{\mathbb{R}}

\newcommand{\nn}{\nonumber}

\newcommand{\spa}{\ , \ \ }

\newcommand{\ds}{\displaystyle}

\newcommand{\tr}{\mathop{{\rm Tr}}}

\newcommand{\ads}{\mbox{AdS}}

\begin{document}

\begin{titlepage}

\ \
 \vskip 2 cm

\centerline{\LARGE \bf The $SU(2)\times SU(2)$ sector in the string
dual of } \vskip 0.2cm \centerline{\LARGE \bf $\CN=6$ superconformal
Chern-Simons theory} \vskip 1.7cm

\centerline{\large {\bf Gianluca Grignani$\,^{1}$}, {\bf Troels
Harmark$\,^{2}$} and {\bf Marta Orselli$\,^{2}$} }

\vskip 0.5cm

\begin{center}
\sl $^1$ Dipartimento di Fisica, Universit\`a di Perugia,\\
I.N.F.N. Sezione di Perugia,\\
Via Pascoli, I-06123 Perugia, Italy\\
\vskip 0.4cm
\sl $^2$ The Niels Bohr Institute  \\
\sl  Blegdamsvej 17, 2100 Copenhagen \O , Denmark \\
\end{center}
\vskip 0.5cm

\centerline{\small\tt grignani@pg.infn.it, harmark@nbi.dk,
orselli@nbi.dk}

\vskip 1.5cm

\centerline{\bf Abstract} \vskip 0.2cm \noindent We examine the string dual of the recently
constructed $\CN=6$ superconformal Chern-Simons theory of Aharony, Bergman, Jafferis and Maldacena
(ABJM theory). We focus in particular on the $SU(2)\times SU(2)$ sector. We find a sigma-model limit
in which the resulting sigma-model is two Landau-Lifshitz models added together. We consider a
Penrose limit for which we can approach the $SU(2)\times SU(2)$ sector. Finally, we find a new Giant
Magnon solution in the $SU(2)\times SU(2)$ sector corresponding to one magnon in each $SU(2)$. We put
these results together to find the full magnon dispersion relation and we compare this to recently
found results for ABJM theory at weak coupling.


\end{titlepage}

\pagestyle{plain} \setcounter{page}{1}

\tableofcontents

\section{Introduction and summary}
\label{sec:intro}

For the last decade, the duality between $\CN=4$ superconformal
Yang-Mills (SYM) theory and type IIB string theory on $\ads_5\times
S^5$ have been celebrated as the one example of an exact duality
between gauge theory and string theory. Recently, developments,
initiated by Bagger, Lambert and Gustavsson \cite{Bagger:2006sk}, in
finding the superconformal world-volume theory for multiple
M2-branes led Aharony, Bergman, Jafferis and Maldacena to construct
a new $\CN=6$ superconformal Chern-Simons theory (ABJM theory)
\cite{Aharony:2008ug} which should be the world-volume theory of
multiple M2-brane on $\C^4 /\Z_k$. Based on this they conjectured a
new duality between ABJM theory and type IIA string theory on
$\ads_4\times \C P^3$. This is a new exact duality
between gauge theory and string theory.%
\footnote{The construction of the $\CN = 6$ superconformal Chern-Simons theory is based on
\cite{Gaiotto:2007qi}. For papers considering the Bagger-Lambert-Gustavsson theory
see~\cite{Gustavsson:2008dy}. For papers considering the ABJM theory see
\cite{Benna:2008zy,Ezhuthachan:2008ch,Nishioka:2008gz,Minahan:2008hf}.}

The ABJM theory consists of two Chern-Simons theories of level $k$
and $-k$ and each with gauge group $SU(N)$, which means that the
total gauge symmetry is $SU(N)\times SU(N)$. In addition it has two
pairs of chiral superfields transforming in the bifundamental
representations of $SU(N) \times SU(N)$. The R-symmetry is $SU(4)$
in accordance with the $\CN=6$ supersymmetry of the theory. It was
observed in \cite{Aharony:2008ug} that one can define a 't Hooft
coupling $\lambda = N/k$ and that in the 't Hooft limit
$N\rightarrow \infty$ with $\lambda$ fixed one has a continuous
coupling $\lambda$ and that the ABJM theory is weakly coupled for
$\lambda \ll 1$. The ABJM theory is conjectured to be dual to
M-theory on $\ads_4\times S^7 / \Z_k$ with $N$ units of four-form
flux. In the limit of large $k$ one has roughly speaking that $S^7 /
\Z_k \simeq \C P^3 \times S^1$ which thus means that ABJM theory in
the 't Hooft limit is dual to type IIA string theory on
$\ads_4\times \C P^3$. This duality is valid for $\lambda \gg 1$ and
the type IIA string description holds when $k \gg N^{1/5}$.

Having this new $\ads_4 / \mbox{CFT}_3$ duality naturally brings up
the question of how similar it is with the $\ads_5 / \mbox{CFT}_4$
duality. We see that despite the fact that $k$ is integer valued we
can still define a continuous 't Hooft coupling and we have a
weak/strong duality between the ABJM theory and type IIA string
theory. Furthermore, Minahan and Zarembo \cite{Minahan:2008hf} have recently
provided evidence that ABJM theory is integrable to second order in
$\lambda$ by finding an integrable $SU(4)$ spin chain. This thus
brings the hope that ABJM theory is integrable, just as has been
seen in the case of $\CN=4$ SYM theory \cite{Minahan:2002ve}. However, there is
one notable difference between the  $\ads_4 / \mbox{CFT}_3$ and
$\ads_5 / \mbox{CFT}_4$ dualities, namely that while one has the
maximal number of 32 supercharges in the $\ads_5 / \mbox{CFT}_4$
case, the number of supercharges in the $\ads_4 / \mbox{CFT}_3$
duality is 24. This means that it can be more challenging to
interpolate from weak to strong coupling in the $\ads_4 /
\mbox{CFT}_3$ duality.

In this paper we study further the question of integrability in the
new $\ads_4 / \mbox{CFT}_3$ duality. We do this by investigating the
$SU(2)\times SU(2)$ sector of the ABJM theory on the string side.
For $\lambda\ll 1$ Minahan and Zarembo found that there is a
decoupled $SU(2)\times SU(2)$ sector in the $SU(4)$ spin chain
\cite{Minahan:2008hf}. In this sector the spin chain Hamiltonian is
that of two $XXX_{1/2}$ Heisenberg spin chains.

We find on the string side a limit of type IIA string theory on
$\ads_4\times \C P^3$ that corresponds to the $SU(2)\times SU(2)$
sector. In this limit the string sigma-model becomes that of two
Landau-Lifshitz models, thus in accordance with the results of
\cite{Minahan:2008hf}. As one might expect, this means that the
S-matrices matches up to second-order corrections for small momenta.
We also find a dispersion relation of the form
\begin{equation}
\Delta = \frac{1}{2} + \frac{\lambda}{2} p^2
\end{equation}
This dispersion relation holds in the limit of $p \rightarrow 0$
with large but fixed $\lambda$. However, it does not match the one
found by Minahan and Zarembo in \cite{Minahan:2008hf}.

To examine further the dispersion relation on the string theory side
we consider a Penrose limit corresponding to the $SU(2)\times SU(2)$
sector (see \cite{Nishioka:2008gz} for another Penrose limit dual to ABJM theory).
We find in particular the dispersion relation for an $SU(2)\times
SU(2)$ magnon
\begin{equation}
\Delta = \sqrt{ \frac{1}{4} + \frac{\lambda}{2} p^2 }
\end{equation}
This holds for $p\rightarrow 0$ with $\lambda p^2$ fixed. This
result is consistent with our sigma-model limit and is furthermore consistent with the Penrose limit of \cite{Nishioka:2008gz}.

We find moreover a new Giant Magnon solution in the $SU(2)\times SU(2)$ sector of type IIA string
theory on $\ads_4 \times \C P^3$, following the Giant Magnon solutions in $\ads_5\times S^5$
\cite{Hofman:2006xt,Arutyunov:2006gs}. The Giant Magnon solution in the $SU(2)\times SU(2)$ sector
that we find has the interesting feature that it consists of two Giant Magnons, one for each $SU(2)$.
As for the Hofman-Maldacena Giant Magnon solution on $\ads_5\times S^5$, this
is a closed string solution with open boundary conditions in two azimuthal directions.%
\footnote{It would be
interesting to see if by considering an orbifold of $\C P^3$~\cite{Benna:2008zy} it would be possible
to identify the string endpoints to make of this a legitimiate closed string solution, as was done
in~\cite{Astolfi:2007uz,Ramadanovic:2008qd} for the $\ads_5\times S^5$ Giant Magnon.}

 From our new
Giant Magnon solution we get the following result for the dispersion relation (for a single magnon)
\begin{equation}
\Delta = \sqrt{ 2 \lambda } \left| \sin  \frac{p}{2}  \right|
\end{equation}
which holds for $\lambda \rightarrow \infty$ and fixed $p$. This
result is consistent with the Penrose limit result.

Combining our results from the sigma-model limit, the Penrose limit
and the Giant Magnon analysis, we find the dispersion relation
\begin{equation}
\label{dispus} \Delta = \sqrt{ \frac{1}{4} + 2 \lambda \sin^2 \Big(
\frac{p}{2} \Big) }
\end{equation}
for $\lambda \gg 1$. For $\lambda \ll 1$ the following dispersion
relation has instead been found \cite{Minahan:2008hf}
\begin{equation}
\label{dispmz} \Delta = \frac{1}{2} + 4 \lambda^2 \sin^2 \Big(
\frac{p}{2} \Big)
\end{equation}
It is evident that \eqref{dispus} and \eqref{dispmz} cannot match,
as one clearly can see in the limit of small momenta.

For the analogous question in the $\ads_5 / \mbox{CFT}_4$ duality it
was found by Beisert that the form of the magnon dispersion relation
is fixed up to a function depending only of the 't Hooft coupling
\cite{Beisert:2005tm}. Assuming that this symmetry argument can be
generalized to the $\ads_4 / \mbox{CFT}_3$ duality, this leads to the
proposal that the magnon dispersion relation  in the $SU(2)\times
SU(2)$ sector for any value of $\lambda$ is of the form
\begin{equation}
\label{dispgen} \Delta = \sqrt{ \frac{1}{4} + h(\lambda) \sin^2
\Big( \frac{p}{2} \Big) }
\end{equation}
where $h(\lambda)$ is a function of $\lambda$. Then our
computations, together with \eqref{dispmz}, shows that
\begin{equation}
\label{hl}
h(\lambda) = \left\{ \begin{array}{c} \ds 4 \lambda^2 + \CO ( \lambda^4 ) \ \mbox{for} \ \lambda \ll 1 \\[4mm] \ds 2 \lambda + \CO ( \sqrt{\lambda} ) \ \mbox{for} \ \lambda \gg 1 \end{array} \right.
\end{equation}
Thus, $h(\lambda)$ is a non-trivial function of the coupling.
This is in contrast with the $\ads_5 / \mbox{CFT}_4$ duality where the same dispersion relation holds for
weak and strong coupling. We believe that this difference is due to the lower amount of supersymmetry of
the $\ads_4 / \mbox{CFT}_3$ duality which indeed makes it more challenging to connect the two sides of the duality.%
\footnote{See \cite{Lin:2005nh} for another case where the
dispersion relation depends non-trivially on the coupling.}

\ \newline \noindent {\bf Note added:} After completing this paper,
Ref.~\cite{Gaiotto:2008cg} appeared on the arXive. This paper has
substantial overlap with our sections \ref{sec:penrose} and
\ref{sec:giant}.

\section{ABJM theory, its spin chain description and its string dual}
\label{sec:ABJM}

The ABJM theory, which is an $\CN=6$ $SU(N)\times SU(N)$
superconformal Chern-Simons theory at level $k$, has two pairs of
chiral superfields, each transforming in a bifundamental
representation of $SU(N)\times SU(N)$. The theory has an explicit
$SU(2) \times SU(2)$ R-symmetry with one pair of superfields being
in the spin $1/2$ representation of the first $SU(2)$ and the other
pair in the second $SU(2)$. Furthermore, the R-symmetry of the
theory has been shown to be enhanced to $SU(4)$ (further enhanced to
$SO(8)$ for $k=1,2$).

ABJM introduced a 't Hooft coupling $\lambda = N/k$. In the 't Hooft
limit $N \rightarrow \infty$ with $\lambda$ fixed, $\lambda$ is a
continuous parameter. For $\lambda \ll 1$ the ABJM theory is weakly
coupled.

We consider the ABJM theory on $\R \times S^2$, thus the global
bosonic symmetry group is $SO(2,3) \times SU(4)$. By the
state/operator correspondence a state for the theory on $\R \times
S^2$ is mapped to an operator for the theory on $\R^3$ with the
scaling dimension $\Delta$ given by the energy in units of the
two-sphere radius.

Focusing on the scalars in the theory we have a pair of complex
scalars $A_1,A_2$ which transform in the $N \times \bar{N}$
representation of $SU(N)\times SU(N)$ and a pair of complex scalars
$B_1,B_2$ which transform in the $\bar{N}\times N$ representation.
One can group these scalars into multiplets of the R-symmetry group
$SU(4)$
\begin{equation}
\label{scalars} Z^a = (A_1,A_2,B_1^\dagger,B_2^\dagger) \spa
Z_a^\dagger = (A^\dagger_1,A^\dagger_2,B_1,B_2)
\end{equation}
with $Z^a$ transforming in the fundamental representation and
$Z^\dagger_a$ in the anti-fundamental representation of $SU(4)$. All
scalars have conformal dimension $\Delta = 1/2$ and transform in the
trivial representation of the $SO(3)$ symmetry.

We have in addition a covariant derivative $D_\mu$ transforming in
the spin 1 representation of $SO(3)$ and in the trivial
representation of $SU(4)$. The scaling dimension is $\Delta = 1$. We
write the three components as $D_-$, $D_0$ and $D_+$ according to
the Cartan generator $S$ of $SO(3)$ ($i.e.$ with eigenvalues $-1$,
$0$ and $1$).

The fermions of the ABJM theory are the superpartners of the
scalars, thus they transform in the fundamental and anti-fundamental
representations of $SU(4)$, and they transform in the spin $1/2$
representation of the $SO(3)$ symmetry.

\subsubsection*{Scalar operators and the $SU(4)$ spin chain}

If we wish to construct gauge-invariant single-trace operators only
from scalars we see that this should be done by alternatingly
combining the scalars $Z^a$ with the scalars $Z^\dagger_a$ since
then we can contract the indices with respect to the $SU(N)\times
SU(N)$ gauge group. Thus, we can consider single-trace operators of
the form \cite{Minahan:2008hf}\footnote{These operators resemble
scalar operators in the $\CN=2$ superconformal Quiver Gauge Theories
\cite{Bertolini:2002nr,Mukhi:2002ck}.}
\begin{equation}
\label{scalarop} \CO = W^{b_1 b_2 \cdots b_n}_{a_1 a_2 \cdots a_n}
\tr ( Z^{a_1} Z^\dagger_{b_1} \cdots Z^{a_n} Z^\dagger_{b_n} )
\end{equation}
In \cite{Minahan:2008hf} the two-loop dilatation operator was
considered for this class of operators interpreting the operator
\eqref{scalarop} as a spin chain of length $2n$ with the spins in
the odd sites transforming in the fundamental and the spins in the
even sites in the anti-fundamental representations of $SU(4)$. This
is in analogy with the analysis of the scalar operators of $\CN=4$
SYM \cite{Minahan:2002ve}. The result is the anomalous dimension
\cite{Minahan:2008hf}
\begin{equation}
\label{MZdelta} \Delta = \Delta_0 + \frac{\lambda^2}{2}
\sum_{l=1}^{2n} ( 2 - 2 P_{l,l+2} + P_{l,l+2} K_{l,l+1} + K_{l,l+1}
P_{l,l+2} )
\end{equation}
with $P$ being the permutation operator and $K$ the trace operator.

Amazingly, it was shown in \cite{Minahan:2008hf} that
\eqref{MZdelta} is integrable, thus suggesting that ABJM theory in
the 't Hooft limit has an integrable structure in analogy with that
of $\CN=4$ SYM. This indeed makes it a very interesting theory to
study. The explicit Bethe equations and dispersion relation for the
integrable $SU(4)$ spin chain are written down in
\cite{Minahan:2008hf}.

\subsubsection*{The $\ads_4 / \mbox{CFT}_3$ duality}

The ABJM theory is conjectured to be the world-volume theory on $N'
= N k $ coincident M2-branes on the orbifold $\C^4 / \Z_k$
\cite{Aharony:2008ug}. Taking the near-horizon limit of the geometry of $N'$
M2-branes on $\C^4 / \Z_k$ gives the $\ads_4\times S^7 / \Z_k $
geometry
\begin{equation}
ds_{11}^2 = \frac{\hat{R}^2}{4} \Big( - \cosh^2 \rho dt^2 + d\rho^2
+ \sinh^2 \rho d\hat{\Omega}_2^2 \Big) + \hat{R}^2 ds_{S^7 /\Z_k }^2
\end{equation}
with $\hat{R}^2 = (2^5 \pi^2 N' )^{1/3} l_p^2$ and with the four
form field strength
\begin{equation}
F_{(4)} = \frac{3 \hat{R}^3}{8} \epsilon_{\rm \ads_4}
\end{equation}
where $\epsilon_{\rm \ads_4}$ is the unit volume form on $\ads_4$.
We can parameterize the $S^7 / \Z_k$ geometry using the four complex
scalars $z_1,z_2,z_3,z_4$ such that
\begin{equation}
ds_{S^7 /\Z_k }^2 = \sum_{a=1}^4 dz_a d\bar{z}_a \spa \sum_{a=1}^4
z_a \bar{z}_a = 1
\end{equation}
The orbifolding is implemented as follows. We write
\begin{equation}
z_a = \mu_a e^{i \phi_a}
\end{equation}
Then we span an $S^7$ if $\sum_{a=1}^4 \mu_a^2 = 1$. To each angle
$\phi_a$ we associate the angular momentum
\begin{equation}
J_a = - i \partial_{\phi_a}
\end{equation}
Write now the angles as
\begin{equation}
\begin{array}{c} \ds \phi_1 = \gamma + \frac{1}{2} ( - \eta_1 - \eta_2 - \eta_3 )
\spa \phi_2 = \gamma + \frac{1}{2} ( \eta_1 + \eta_2 - \eta_3 )
\\[4mm] \ds \phi_3 = \gamma + \frac{1}{2} ( \eta_1 - \eta_2 + \eta_3 )
\spa \phi_4 = \gamma + \frac{1}{2} ( - \eta_1 + \eta_2 + \eta_3
)\end{array}
\end{equation}
The orbifold $S^7 / \Z_k$ is now implemented as the identification
\begin{equation}
\label{gammaid}
\gamma \equiv \gamma + \frac{2\pi}{k}
\end{equation}
We have that
\begin{equation}
J_1+J_2+J_3+J_4 = - i \partial_\gamma
\end{equation}
Thus, we see that the orbifolding is equivalent to the quantization
condition
\begin{equation}
J_1+J_2+J_3+J_4 \in k \Z
\end{equation}
Introducing the three charges
\begin{equation}
R_j = - i \partial_{\eta_j}
\end{equation}
we see that $R_1,R_2,R_3$ are the three Cartan generators for the
$SU(4)$ subgroup of $SO(8)$ which is dual to the $SU(4)$ R-symmetry
of the ABJM theory. In detail,
\begin{equation}
\label{Rchar} R_1 = \frac{1}{2} ( J_1-J_2-J_3+J_4) \spa R_2 =
\frac{1}{2} ( -J_1+J_2-J_3+J_4) \spa R_3 = \frac{1}{2} (
-J_1-J_2+J_3+J_4)
\end{equation}

We can identify the four complex scalars $z_a$ with the four scalar
fields $Z^a$ of the ABJM theory given in \eqref{scalars}. In
particular we see that $Z^a$ transforms in the fundamental
representation with highest weight $(1/2,1/2,1/2)$ in terms of
$(R_1,R_2,R_3)$ while $Z^\dagger_a$ transforms in the
$(1/2,1/2,-1/2)$ anti-fundamental representation.

Write now
\begin{equation}
ds_{S^7 /\Z_k }^2 = ds_{\C P^3}^2 + (d\gamma + A )^2
\end{equation}
Thus the eleven-dimensional metric is
\begin{equation}
ds_{11}^2 = \frac{\hat{R}^2}{4} \Big( - \cosh^2 \rho dt^2 + d\rho^2
+ \sinh^2 \rho d\hat{\Omega}_2^2 \Big) + \hat{R}^2 ds_{\C P^3}^2 +
\hat{R}^2 (d\gamma + A )^2
\end{equation}
Using the standard relation between the M-theory metric and the type
IIA metric, along with the relation $l_p^3 = g_s l_s^3$ and that the
eleven-dimensional radius is $R_{11} = g_s l_s$, we get the
following background of type IIA supergravity given by the metric
\begin{equation}
\label{iiamet}
ds^2 = \frac{R^2}{4} \Big( - \cosh^2 \rho dt^2 + d\rho^2 + \sinh^2
\rho d\hat{\Omega}_2^2 \Big) + R^2 ds_{\C P^3}^2
\end{equation}
with
\begin{equation}
\label{Rdef} \frac{R^2}{l_s^2} = \frac{\sqrt{2^5 \pi^2 N'}}{k}  =
\sqrt{ \frac{2^5 \pi^2 N}{k} } = \sqrt{2^5 \pi^2 \lambda}
\end{equation}
and moreover given by the string coupling constant
\begin{equation}
g_s = \frac{( 2^5 \pi^2 N' )^{1/4}}{k^{3/2}} = \Big( \frac{2^5 \pi^2
N}{k^5} \Big)^{\frac{1}{4}}
\end{equation}
the Ramond-Ramond (RR) four-form field strength
\begin{equation}
F_{(4)} = \frac{3 R^3}{8} \epsilon_{\rm \ads_4}
\end{equation}
and with $A$ being a one-form RR potential corresponding to the
two-form RR field strength $F_{(2)} = d A$. From demanding a small
curvature and a small string coupling one finds that this background
is a valid background for type IIA string theory when $\lambda \gg
1$ and $N \ll k^5$ \cite{Aharony:2008ug}.

When considering the type IIA description we should clearly require
that the dependence on $\gamma$ is absent, we get therefore that we
should only consider operators obeying
\begin{equation}
\label{Jcon} J_1+J_2+J_3+J_4 = 0
\end{equation}
This is in accordance with the construction of single-trace scalars
operators \eqref{scalarop} in the ABJM theory since we see that
these operators indeed obey \eqref{Jcon}.

Note that for fixed $\rho \gg 1$ the $\ads_4$ part of the metric \eqref{iiamet} approaches $\R\times S^2$ as $e^{2\rho} R^2/4 (-dt^2 + d\hat{\Omega}_2^2 )$. Since the conformal dimension $\Delta$ in ABJM theory is the energy in units of the two-sphere radius, we see that we should identify $\Delta$ with
\begin{equation}
\Delta = i \partial_t
\end{equation}

\section{Subsectors of the ABJM theory}
\label{sec:sectors}

In this section we consider decoupled subsectors in the ABJM theory.
A straightforward method to analyze this was provided in
\cite{Harmark:2007px} for $\CN=4$ SYM (for a method based on group
theory see \cite{Beisert:2003jj}). For ABJM theory we should
consider the possible inequalities of the form
\begin{equation}
\Delta_0 \geq m_1 R_1 + m_2 R_2 + m_3 R_3 + m_4 S
\end{equation}
where $\Delta_0$ is the bare scaling operator, $R_j$ are the three
Cartan generators of the $SU(4)$ R-symmetry, $S$ is the Cartan
generator of the $SO(3)$ symmetry and $m_i$ are rational numbers.
Alternatively using \eqref{Rchar} we can express this as
\begin{equation}
\label{genin}
\Delta_0 \geq n_1 J_1 + n_2 J_2 + n_3 J_3 + n_4 J_4 + n_5 S
\end{equation}
assuming the extra restriction \eqref{Jcon} and where $n_i$ are
rational numbers. The upshot is that if the inequality is saturated
for certain operators then those operators comprise a decoupled
sector for the leading contribution to the anomalous dimension
operator $\Delta - \Delta_0$.

\subsubsection*{The $SU(2)\times SU(2)$ sector}

Consider the inequality
\begin{equation}
\Delta_0 \geq J_1 + J_2
\end{equation}
The operators in the ABJM theory that saturate this inequality,
$i.e.$ for which $\Delta_0 = J_1+J_2$, are the ones made out of the
scalars $A_{1,2}$ and $B_{1,2}$. The single-trace operators are thus
of the form
\begin{equation}
\label{su2op} \CO = W^{j_1 j_2 \cdots j_J}_{i_1 i_2 \cdots i_J}
\tr ( A_{i_1} B_{j_1} \cdots A_{i_J} B_{j_J} )
\end{equation}
This constitutes an $SU(2) \times SU(2)$ sector of the ABJM theory,
as found in \cite{Minahan:2008hf}, since the $A_{1,2}$ and $B_{1,2}$
scalars transform in two separate $SU(2)$ subgroups of the $SU(4)$.
From the result \eqref{MZdelta} of \cite{Minahan:2008hf} we see
furthermore that
\begin{equation}
\label{SU2delta} \Delta - J = \lambda^2 \sum_{l=1}^{2J} ( 1 -
P_{l,l+2}  ) = \lambda^2 \sum_{l=1}^J ( 1 - P_{2l-1,2l+1} + 1 -
P_{2l,2l+2} )
\end{equation}
We defined here $J = J_1+J_2 = - J_3 - J_4$. We see that \eqref{SU2delta} corresponds to two decoupled ferromagnetic $XXX_{1/2}$ Heisenberg spin chains, one living at the odd sites and the other at the even sites \cite{Minahan:2008hf}. The spectrum is determined by the following dispersion relation, Bethe equations and momentum constraint
\begin{equation}
\label{dispsu2} \Delta - J = 4 \lambda^2 \left[ \sum_{i=1}^{M_1}
\sin^2 \Big( \frac{p^{(1)}_i}{2} \Big) + \sum_{i=1}^{M_2} \sin^2
\Big( \frac{p^{(2)}_i}{2} \Big) \right]
\end{equation}
\begin{equation}
e^{ip^{(a)}_k J } = \prod_{j=1,j \neq k}^{M_a} S(p^{(a)}_k,p^{(a)}_j) \spa
\sum_{i=1}^{M_1} p^{(1)}_i + \sum_{i=1}^{M_2} p^{(2)}_i = 0
\end{equation}
for $a=1,2$, with the S-matrix given by
\begin{equation}
\label{MZS} S(p_k,p_j) = -
\frac{1+e^{i(p_k+p_j)}-2e^{ip_k}}{1+e^{i(p_k+p_j)}-2e^{ip_j}}
\end{equation}
We see that the two chains affect each other through the momentum
constraint which means that the spectrum is not just given by adding
together two independent Heisenberg spin chains. We also note that
we can infer from \eqref{dispsu2} that the magnon dispersion
relation in the $SU(2)\times SU(2)$ sector is given by
\eqref{dispmz} which in turn reveals that $h(\lambda) = 4\lambda^2 $
for small $\lambda$ in the general dispersion relation
\eqref{dispgen}.

\subsubsection*{Other sectors}

Consider the inequality
\begin{equation}
\Delta_0 \geq J_1+J_2+J_3
\end{equation}
We see that the only operators that can saturate this inequality are
those that have $B_2$ on the even sites and $(Z^1,Z^2,Z^3) =
(A_1,A_2,B_1^\dagger)$ on the odd sites. Thus, we can consider
single-trace operators of the form
\begin{equation}
\CO = W_{a_1 a_2 \cdots a_n} \tr ( Z^{a_1} B_2 \cdots Z^{a_n} B_2 )
\end{equation}
with $a_j = 1,2,3$. This is the $SU(3)$ sector found in \cite{Minahan:2008hf}.

It is furthermore interesting to consider sectors with derivatives. We can only get derivatives in the inequality \eqref{genin} if $n_5 \in \{ -1 , 1 \}$. Consider the inequality
\begin{equation}
\Delta_0 \geq S + J_1+J_2
\end{equation}
For this case we see that at odd sites we can either have $D_+^n A_{1,2}$ or $D_+^n \chi_{A_{1,2}}$ where $\chi_{A_{1,2}}$ is the component of the superpartner of $A_{1,2}$ with $S=1/2$. For even sites we can either have $D_+^n B_{1,2}$ and $D_+^n \chi_{B_{1,2}}$ where $\chi_{B_{1,2}}$ is the component of the superpartner of $B_{1,2}$ with $S=1/2$. This sector generalizes the $SU(2)\times SU(2)$ sector to include both the derivative $D_+$ and a superpartner. This sector could be relevant for studying the cusp anomaly in the ABJM theory.

We can also generalize the $SU(3)$ sector inequality to
\begin{equation}
\Delta_0 \geq S + J_1+J_2+J_3
\end{equation}
At odd sites we have $D_+^n Z^{1,2,3}$ and $D_+^n \chi_{Z^{1,2,3}}$
with $\chi_{Z^{1,2,3}}$ being the superpartner of $Z^{1,2,3}$ with
$S=1/2$, while at even sites we have $D_+^n B_{2}$ and $D_+^n
\chi_{B_{2}}$ with $\chi_{B_{2}}$ being the superpartner of $B_{2}$
with $S=1/2$.

\section{The $SU(2) \times SU(2)$ sigma-model limit}
\label{sec:smlimit}

In this section we take a limit of the type IIA string theory
sigma-model on $\ads_4\times \C P^3$ corresponding to zooming in to
the $SU(2)\times SU(2)$ sector. The idea is that by taking a limit
where $\Delta - J_1 - J_2$ goes to zero, then only the string states
of the $SU(2)\times SU(2)$ sector can survive. This corresponds to a
limit of small momenta, and the leading contribution to $\Delta -
J_1 - J_2$ gives a sigma-model describing the small momentum regime
of the strings in the $SU(2)\times SU(2)$ sector. This type of limit
was first found in \cite{Kruczenski:2003gt} (see also
\cite{Kruczenski:2004kw,Harmark:2008gm}).

In order to understand how to zoom in to the relevant part of the
geometry of $\ads_4\times \C P^3$ we first take a step back and
consider the M-theory background $\ads_4 \times S^7$ corresponding
to M2-branes on $\C^4$. As is clear from Section \ref{sec:ABJM}, the
two $SU(2)$'s are gotten from splitting up $\C^4 = \C^2 \times
\C^2$. In detail the first $SU(2)$ corresponding to $A_{1,2}$ is
then associated to $z_{1,2}$ while the second $SU(2)$ corresponding
to $B_{1,2}$ is associated to $\bar{z}_{3,4}$. We therefore split up
the $S^7$ into two $S^3$'s, one for each $\C^2$, as follows
\begin{equation}
ds_{S^7}^2 = d\theta^2 + \cos^2 \theta d\Omega_3^2 + \sin^2 \theta d{\Omega_3'}^2
\end{equation}
We parameterize the two three-spheres as
\begin{equation}
d\Omega_3^2 = d\psi_1^2 + \sin^2 \psi_1 d\phi_1^2 + \cos^2 \psi_1
d\phi_2^2 \spa d{\Omega_3'}^2 = d\psi_2^2 + \sin^2 \psi_2 d\phi_3^2
+ \cos^2 \psi_2 d\phi_4^2
\end{equation}
with $\phi_a$ being the angles introduced in Section \ref{sec:ABJM}. Introduce now the angles
\begin{equation}
\begin{array}{c} \ds
\theta_1 = 2\psi_1 - \frac{\pi}{2} \spa \theta_2 = 2\psi_2 - \frac{\pi}{2} \spa \varphi_1 = \phi_1-\phi_2 \spa \varphi_2 = \phi_4-\phi_3 \\[3mm] \ds \gamma = \frac{1}{4} ( \phi_1+\phi_2+\phi_3+\phi_4 ) \spa \delta = \frac{1}{4} ( \phi_1+\phi_2-\phi_3-\phi_4 ) \end{array}
\end{equation}
With this, we can write
\begin{equation}
\label{threespheres}
\begin{array}{c} \ds d\Omega_3^2 = \frac{1}{4} d\Omega_2^2 + \Big( d\gamma + d\delta + \frac{1}{2} \sin \theta_1 d\varphi_1 \Big)^2 \spa d\Omega_2^2 = d\theta_1^2 + \cos^2 \theta_1 d\varphi_1^2 \\[3mm] \ds d{\Omega_3'}^2 = \frac{1}{4} d{\Omega_2'}^2 + \Big( d\gamma - d\delta - \frac{1}{2} \sin \theta_2 d\varphi_2 \Big)^2 \spa {d\Omega_2'}^2 = d\theta_2^2 + \cos^2 \theta_2 d\varphi_2^2  \end{array}
\end{equation}
We have
\begin{equation}
\begin{array}{c} \ds S^{(1)}_z \equiv \frac{J_1-J_2}{2} = - i \partial_{\varphi_1} \spa S^{(2)}_z \equiv \frac{J_4-J_3}{2} = - i \partial_{\varphi_2} \\[3mm] \ds J_1+J_2+J_3+J_4 = - i\partial_\gamma \spa J_1+J_2-J_3-J_4 = - i \partial_\delta
\end{array}
\end{equation}
We see that the coordinates $(\theta_i,\varphi_i)$, $i=1,2$,
parameterize a pair of two-spheres. These two two-spheres correspond
to the two $SU(2)$'s. Moreover, we note that we chose the opposite
orientation for $\varphi_1$ and $\varphi_2$ in the two $\C^2$'s
since one $SU(2)$ corresponds to $(z_1,z_2)$ ($A_{1,2}$ in the ABJM
theory) while the other $SU(2)$ to $(\bar{z}_3,\bar{z}_4)$
($B_{1,2}$ in the ABJM theory). This gives the two Cartan generators
$S_z^{(i)}$ corresponding to the total spins for the two $SU(2)$'s.

We can now implement the orbifolding of the $S^7$ by the identification \eqref{gammaid}. In order to zoom in to the $SU(2)\times SU(2)$ sector we set
\begin{equation}
\rho=0 \spa \theta= \frac{\pi}{4}
\end{equation}
This can be justified further since in the limit we take below one can check that the transverse excitations in the $\rho$ and $\theta$ directions become infinitely heavy in the limit, just as it happens in the $SU(2)$ sigma-model limit of $\ads_5\times S^5$ \cite{Harmark:2008gm}. We should thus consider the eleven-dimensional metric
\begin{equation}
ds_{11}^2 = - \frac{\hat{R}^2}{4} dt^2 + \frac{\hat{R}^2}{2} (  d\Omega_3^2 +
d{\Omega_3'}^2 )
\end{equation}
with the identification \eqref{gammaid} where $\hat{R}$ is given in Section \ref{sec:ABJM}. We find that
\begin{equation}
ds_{11}^2 = - \frac{\hat{R}^2}{4} dt^2 + \hat{R}^2 (d\gamma+A)^2 + \hat{R}^2 \Big[
\frac{1}{8} d\Omega_2^2 + \frac{1}{8} d{\Omega_2'}^2 + (d\delta +
\omega)^2  \Big]
\end{equation}
with the one-forms $A$ and $\omega$ given by
\begin{equation}
A = \frac{1}{4} ( \sin \theta_1 d\varphi_1 - \sin \theta_2
d\varphi_2) \spa \omega = \frac{1}{4} ( \sin \theta_1 d\varphi_1 +
\sin \theta_2 d\varphi_2)
\end{equation}
The type IIA background then has the ten-dimensional metric
\begin{equation}
\label{theiia} ds^2 = - \frac{R^2}{4} dt^2  + R^2 \Big[ \frac{1}{8}
d\Omega_2^2 + \frac{1}{8} d{\Omega_2'}^2 + (d\delta + \omega)^2
\Big]
\end{equation}
with $R$ given in \eqref{Rdef}.

As explained in Section \ref{sec:sectors} the $SU(2)\times SU(2)$ sector is obtained by considering states for which $\Delta - J_1 - J_2$ is small. To implement this as a sigma-model limit we make the coordinate transformation
\begin{equation}
\label{tt} \tilde{t} = \frac{1}{J^2} t \spa \chi = \delta -
\frac{1}{2} t
\end{equation}
so that
\begin{equation}
\label{tilH}
\tilde{H} \equiv - i \partial_{\tilde{t}} = J^2 \Big( \Delta - \frac{1}{2} ( J_1+J_2-J_3-J_4) \Big) \spa J_1+J_2-J_3-J_4 = - i\partial_\chi
\end{equation}
where we have defined
\begin{equation}
J \equiv J_1+J_2
\end{equation}
Using the condition \eqref{Jcon} we see that \eqref{tilH} can be
written as
\begin{equation}
\tilde{H} \equiv i \partial_{\tilde{t}} = J^2 ( \Delta - J ) \spa 2J
= - i\partial_\chi
\end{equation}
We see here that taking $J \rightarrow \infty$ corresponds to
zooming in to the regime where $\Delta-J$ is of order $1/J^2$. This
corresponds to the energy scale in which we see the individual
magnon states in the spin chain description. We see from \eqref{tt}
that we are zooming in close to $\delta = t/2$. This is a
null-geodesic in the metric \eqref{theiia}. This null-geodesic
corresponds to a chiral primary of the ABJM theory with $\Delta =
J$.

Employing the coordinate transformation \eqref{tt} we get the type
IIA metric
\begin{equation}
ds^2 = R^2 \left[ ( J^2 d\tilde{t} + d\chi+\omega ) ( d\chi  +
\omega )  +  \frac{1}{8} d\Omega_2^2 + \frac{1}{8} d{\Omega_2'}^2
\right]
\end{equation}
Consider now the bosonic sigma-model Lagrangian
\begin{equation}
\label{lagr} \CL = - \frac{1}{2} G_{\mu \nu} h^{\alpha \beta}
\partial_\alpha x^{\mu} \partial_\beta x^{\nu}
\end{equation}
We pick the gauge
\begin{equation}
\label{gaugec} \tilde{t} = \kappa \tau \spa p_\chi = \mbox{const.}
\spa h_{\alpha\beta} = \eta_{\alpha\beta}
\end{equation}
with $2\pi l_s^2 p_\chi = \partial \CL /
\partial
\partial_\tau \chi$. The Lagrangian \eqref{lagr} is then found to be
\begin{equation}
\label{lagr2} \frac{2}{R^2} \CL = (\kappa J^2 + \partial_\tau \chi +
\omega_\tau) (\partial_\tau \chi + \omega_\tau) - (\chi' +
\omega_\sigma)^2 + \frac{1}{8} \sum_{i=1}^2 \Big[ (\partial_\tau
\theta_i)^2 - {\theta_i'}^2 + \cos^2 \theta_i [ (\partial_\tau
\varphi_i)^2  - {\varphi_i'}^2 ] \Big]
\end{equation}
with $\omega = \omega_\tau d\tau + \omega_\sigma d\sigma$ and with
prime denoting the derivative with respect to $\sigma$. The Virasoro
constraints are
\begin{equation}
\label{conn}
\begin{array}{c} \ds
( \kappa J^2 + \partial_\tau \chi + \omega_\tau ) ( \chi' +
\omega_\sigma )  + \frac{1}{8} \sum_{i=1}^2 \Big[ \partial_\tau
\theta_i \theta_i' + \cos^2 \theta_i
\partial_\tau \varphi_i \varphi_i' \Big] = 0
\\[4mm] \ds
( \kappa J^2 + \partial_\tau \chi + \omega_\tau) (\partial_\tau \chi
+ \omega_\tau) + (\chi' + \omega_\sigma)^2 + \frac{1}{8}
\sum_{i=1}^2 \Big[ (\partial_\tau \theta_i)^2 + {\theta_i'}^2 +
\cos^2 \theta_i [  (\partial_\tau \varphi_i)^2  +  {\varphi_i'}^2 ]
\Big] = 0 \end{array}
\end{equation}
We have
\begin{equation}
\label{pchi} p_\chi = \frac{R^2}{2\pi l_s^2} \Big( \frac{\kappa
J^2}{2} +
\partial_\tau \chi + \omega_\tau \Big)
\end{equation}
Since $\tilde{t}$ measures the time corresponding to the energy
scale $\tilde{H}$ we should consider the velocities with respect to
$\tilde{t}$ to be finite in the $J\rightarrow \infty$ limit. Hence
for example $\partial_\tau \chi = \kappa \partial_{\tilde{t}} \chi$.
Inserting this in \eqref{pchi}, we see that $\partial_\tau \chi
\rightarrow 0$ and using that $2J = \int_0^{2\pi} p_\chi$ we get
from \eqref{pchi} that
\begin{equation}
\kappa = \frac{4 l_s^2}{J R^2}
\end{equation}
which is seen to go to zero in the $J\rightarrow \infty$ limit.
Taking now the $J\rightarrow \infty$ limit of the Lagrangian
\eqref{lagr2} and the constraints \eqref{conn} we get
\begin{equation}
\begin{array}{c} \ds
\frac{2}{R^2} \CL = \frac{16 l_s^4}{R^4} ( \dot{\chi} +
\omega_{\tilde{t}} ) - (\chi' + \omega_\sigma)^2 - \frac{1}{8}
\sum_{i=1}^2 \Big[
 {\theta_i'}^2 + \cos^2 \theta_i {\varphi_i'}^2 \Big] \\[4mm] \ds
\chi' + \omega_\sigma = 0 \spa  \frac{16 l_s^4}{R^4}  ( \dot{\chi} +
\omega_{\tilde{t}}) +  \frac{1}{8} \sum_{i=1}^2 \Big[ {\theta_i'}^2
+ \cos^2 \theta_i {\varphi_i'}^2 \Big] = 0
\end{array}
\end{equation}
Here the dot denote the derivative with respect to $\tilde{t}$. We see that the constraints fix
$\chi$ in terms of $\theta_i$ and $\varphi_i$. Thus we can eliminate $\chi$ to get the gauge fixed
Lagrangian
\begin{equation}
\frac{2}{R^2} \CL = \frac{16 l_s^4}{R^4} \omega_{\tilde{t}} -
\frac{1}{8} \sum_{i=1}^2 \Big[ {\theta_i'}^2 + \cos^2 \theta_i
{\varphi_i'}^2 \Big]
\end{equation}
From this we finally get the action for the sigma-model model in the
$J\rightarrow \infty$ limit as
\begin{equation}
\label{LLL} I = \frac{J}{4\pi} \sum_{i=1}^2 \int d\tilde{t}
\int_0^{2\pi} d\sigma \Big[ \sin \theta_i \dot{\varphi}_i - \pi^2
\lambda \Big( {\theta_i'}^2 + \cos^2 \theta_i {\varphi_i'}^2 \Big)
\Big]
\end{equation}
This is supplemented with the momentum constraint
\begin{equation}
\label{momm} \sum_{i=1}^2 \int_0^{2\pi} d\sigma \sin \theta_i \varphi_i' =0
\end{equation}

Thus, in conclusion the result of taking the $SU(2)\times SU(2)$
sigma-model limit is that we obtain two Landau-Lifshitz models added
together \eqref{LLL}, one for each $SU(2)$, which only affect each
other through the momentum constraint \eqref{momm}. Since the
Landau-Lifshitz model corresponds to the long wave-length
$J\rightarrow \infty$ limit of the $XXX_{1/2}$ Heisenberg spin chain
our result is consistent with finding two Heisenberg spin chains in
the $SU(2)\times SU(2)$ sector of ABJM theory at $\lambda \ll 1$
\cite{Minahan:2008hf}.

It is interesting to compare further the integrable structure that
we found here on the string side with the integrable structure
\eqref{dispsu2}-\eqref{MZS} found on the weakly coupled ABJM theory.
Using the analysis of \cite{Klose:2006dd} we can write the Bethe
equations and dispersion relation corresponding to
\eqref{LLL}-\eqref{momm} as
\begin{equation}
\label{dispLL} \Delta - J = \frac{\lambda}{2} \left[
\sum_{i=1}^{M_1} (p^{(1)}_i)^2 + \sum_{i=1}^{M_2} (p^{(2)}_i)^2
\right]
\end{equation}
\begin{equation}
e^{ip^{(a)}_k J } = \prod_{j=1,j \neq k}^{M_a}
S(p^{(a)}_k,p^{(a)}_j) \spa \sum_{i=1}^{M_1} p^{(1)}_i +
\sum_{i=1}^{M_2} p^{(2)}_i = 0
\end{equation}
for $a=1,2$, with the S-matrix given by
\begin{equation}
S(p_k,p_j) = \frac{\frac{1}{p_k} - \frac{1}{p_j} + i }{\frac{1}{p_k}
- \frac{1}{p_j} - i }
\end{equation}
Comparing this with \eqref{dispsu2}-\eqref{MZS} found in the weakly
coupled ABJM theory we see that the S-matrices coincide for small
momenta (up to order $p^2$) which again is as expected since the
Landau-Lifshitz model describes the long wave-length expansion of
the Heisenberg spin chain. However, the dispersion relations are
clearly different, if one compares them in the $p\rightarrow 0$
limit. See the introduction in Section \ref{sec:intro} for a
discussion of this point.

\section{The $SU(2) \times SU(2)$ Penrose limit}
\label{sec:penrose}

In this section we consider a Penrose limit of the $\ads_4\times \C
P^3$ background which corresponds to the $SU(2)\times SU(2)$
sigma-model limit of Section \ref{sec:smlimit}, following
\cite{Berenstein:2002jq,Bertolini:2002nr}. Another Penrose limit of
$\ads_4\times \C P^3$ have been considered in
\cite{Nishioka:2008gz}. We comment on the relation between the two
Penrose limits below.

We choose in the following to consider only the bosonic string modes
for simplicity. To take the Penrose limit we consider first the
metric for $\ads_4\times \C P^3$
\begin{equation}
\label{adscp} ds^2 = \frac{R^2}{4} \left( - \cosh^2 \rho dt^2 +
d\rho^2 + \sinh^2 \rho d\hat{\Omega}^2_2 \right) + R^2 ds_{\C P^3}^2
\end{equation}
where the $\C P^3$ metric is
\begin{equation}
\label{cp3} ds_{\C P^3}^2 = d\theta^2 + \frac{\cos^2 \theta}{4}
d\Omega_2^2 + \frac{\sin^2 \theta}{4} d{\Omega_2'}^2 +  4 \cos^2
\theta \sin^2 \theta ( d \delta + \omega )^2
\end{equation}
with
\begin{equation}
\omega = \frac{1}{4} \sin \theta_1 d\varphi_1 + \frac{1}{4} \sin
\theta_2 d\varphi_2
\end{equation}
where we used the angles introduced in Section \ref{sec:smlimit}.
Define
\begin{equation}
t' = t \spa \chi = \delta - \frac{1}{2} t
\end{equation}
In these coordinates the metric \eqref{adscp} takes the form
\begin{align}
\label{adscp2} ds^2 = & - \frac{R^2}{4} {dt'}^2 ( 1- 4\cos^2 \theta \sin^2\theta  +
\sinh^2 \rho
) + \frac{R^2}{4}( d\rho^2 + \sinh^2 \rho d\hat{\Omega}_2^2 ) \nn \\
& + R^2 \left[ d\theta^2 +\frac{\cos^2 \theta}{4} d\Omega_2^2
+\frac{\sin^2 \theta}{4} d{\Omega_2'}^2 + 4 \cos^2
\theta \sin^2 \theta  ( dt' + d \chi
+ \omega )( d \chi + \omega )\right]
\end{align}
We have that
\begin{equation}
\label{dmj}
\Delta - J = i \partial_{t'} \spa 2 J = -i \partial_\chi
\end{equation}
Here $\Delta -J$ is the energy we are interested in measuring for the $SU(2)\times SU(2)$ sector.

Define now the rescaled coordinates
\begin{equation}
\label{penlim} v = R^2 \chi \spa u_4 = R \Big( \theta -
\frac{\pi}{4} \Big) \spa r = \frac{R}{2} \rho \spa x_a = R \varphi_a
\spa y_a = R \theta_a
\end{equation}
with $a=1,2$. Then the Penrose limit $R \rightarrow \infty$ gives the following metric for a type IIA pp-wave background
\begin{equation}
\label{ppmet}
ds^2 = dv dt'  + \sum_{i=1}^4 ( du_i^2 - u_i^2 {dt'}^2 )  + \frac{1}{8} \sum_{a=1}^2 (dx_a^2 + dy_a^2 + 2 dt' y_a dx_a )
\end{equation}
where $r^2 = \sum_{i=1}^3 u_i^2$ and $dr^2 + r^2 d\hat{\Omega}_2^2 = \sum_{i=1}^3 du_i^2$. The RR field strengths for this pp-wave background are given by
\begin{equation}
\label{ppRR}
F_{(2)} = dt' du_4 \spa F_{(4)} = 3 dt' du_1 du_2 du_3
\end{equation}
The type IIA pp-wave background \eqref{ppmet}-\eqref{ppRR} has 24 supercharges
and it is the same background found from another Penrose limit of $\ads_4\times \C P^3 $
in \cite{Nishioka:2008gz} though in another coordinate system.
The background \eqref{ppmet}-\eqref{ppRR} was originally found in \cite{Sugiyama:2002tf,Hyun:2002wu}.

We notice that the pp-wave background \eqref{ppmet}-\eqref{ppRR} has
two flat directions $x_1$ and $x_2$. Thus the coordinate system of
this pp-wave is similar to the one found by a Penrose limit in
\cite{Bertolini:2002nr}. The Penrose limit of
\cite{Bertolini:2002nr} is particularly well suited to consider the
$SU(2)$ sector of $\ads_5\times S^5$, as explained in
\cite{Harmark:2006ta,Harmark:2008gm}. Similarly, we shall see below
that the Penrose limit given by \eqref{penlim} is particularly well
suited for the $SU(2)\times SU(2)$ sector of $\ads_4 \times \C P^3$.

We choose the gauge
\begin{equation}
t'=c\tau~,~~~h_{\alpha,\beta}=\eta_{\alpha,\beta} \label{pengc}
\end{equation}
and we get, for the bosonic fields, the following gauge fixed
Lagrangian
\begin{equation}
 \CL=\frac{1}{2}\sum_{i=1}^4\left[\left(\partial_{\tau}u_i\right)^2
 -u_i'^2-c^2u_i^2\right]+\frac{c}{8}\sum_{a=1}^2y_a\partial_{\tau}x_a
 +\frac{1}{16}\sum_{a=1}^2\left[
\left(\partial_{\tau}x_a\right)^2+\left(\partial_{\tau}y_a\right)^2
 -x_a'^2-y_a'^2\right]
\label{penlag}
\end{equation}
The bosonic light-cone Hamiltonian is then given by
\begin{equation}
cH_{lc}= \frac{1}{2\pi l_s^2} \int_0^{2\pi} d\sigma
\left\{\frac{1}{2}\sum_{i=1}^4\left[\left(\partial_{\tau}u_i\right)^2
 +u_i'^2+c^2u_i^2\right]+\frac{1}{16}\sum_{a=1}^2\left[
\left(\partial_{\tau}x_a\right)^2+\left(\partial_{\tau}y_a\right)^2
 +x_a'^2+y_a'^2\right] \right\}
 \label{penham}
 \end{equation}
The mode expansion for the bosonic fields can be written as
\begin{equation}
u_i (\tau,\sigma ) = \frac{i}{\sqrt{2}} \sum_{n\in \Z}
\frac{1}{\sqrt{\Omega_n}} \Big[ \hat{a}^i_n e^{-i ( \Omega_n \tau -
n \sigma ) } - (\hat{a}^i_n)^\dagger e^{i ( \Omega_n \tau - n \sigma
) } \Big]
\end{equation}
\begin{equation}
\label{zmode} z_a(\tau,\sigma) = 2 \sqrt{2} \, e^{i\frac{ c
\tau}{2}} \sum_{n \in \Z} \frac{1}{\sqrt{\omega_n}} \Big[ a_n^a
e^{-i ( \omega_n \tau - n \sigma ) } -  (\tilde{a}^a)^\dagger_n e^{i
( \omega_n \tau - n \sigma ) } \Big]
\end{equation}
where $\Omega_n=\sqrt{c^2+n^2}$, $\omega_n=\sqrt{\frac{c^2}{4}+n^2}$
and we defined
$z_a(\tau,\sigma)=x_a(\tau,\sigma)+iy_a(\tau,\sigma)$. The canonical
commutation relations $[x_a(\tau,\sigma),p_{x_b}(\tau,\sigma')] =
i\delta_{ab} \delta (\sigma-\sigma')$,
$[y_a(\tau,\sigma),p_{y_b}(\tau,\sigma')] = i\delta_{ab}\delta
(\sigma-\sigma')$ and $[u_i(\tau,\sigma),p_j(\tau,\sigma')] =
i\delta_{ij} \delta (\sigma-\sigma')$ follows from
\begin{equation}
\label{comrel} [a_m^a,(a_n^b)^\dagger] = \delta_{mn} \delta_{ab}\spa
[\tilde{a}_m^a,(\tilde{a}_n^b)^\dagger] = \delta_{mn}
\delta_{ab}\spa [\hat{a}^i_m,(\hat{a}^j_n)^\dagger] = \delta_{mn}
\delta_{ij}
\end{equation}
Employing \eqref{comrel} we obtain the bosonic spectrum
\begin{equation}
c H_{\rm lc} = \sum_{i=1}^4 \sum_{n\in \Z} \sqrt{n^2+c^2}\,
\hat{N}^i_n+\sum_{a=1}^2\sum_{n\in \Z}
\left(\sqrt{\frac{c^2}{4}+n^2} - \frac{c}{2}\right) M_n^a
+\sum_{a=1}^2\sum_{n\in \Z} \left(\sqrt{\frac{c^2}{4}+n^2}+
\frac{c}{2}\right) N_n^a \label{penspectrum}
\end{equation}
with the number operators $\hat{N}^i_n = (\hat{a}^i_n)^\dagger
\hat{a}^i_n$, $M_n^a = (a^a)^\dagger_n a^a_n$ and $N_n^a =
(\tilde{a}^a)^\dagger_n \tilde{a}_n^a$, and with the level-matching
condition
\begin{equation}
\sum_{n\in \Z}n \left[\sum_{i=1}^4  \hat{N}^i_n+\sum_{a=1}^2
\left(M_n^a + N_n^a\right)\right]
 = 0
\end{equation}

The constant $c$ can be fixed from the term
$\frac{c}{2}\partial_{\tau}v$ in the full Lagrangian. In fact we
have that $2\pi l_s^2p_v=\partial \CL/\partial \partial_{\tau} v$
which gives
\begin{equation}
c=\frac{4l_s^2J}{R^2}=\frac{J}{\pi\sqrt{2\lambda}} \label{cconstant}
\end{equation}
where we again used that $\int_0^{2\pi}d\sigma p_\chi=2J$. Using
\eqref{cconstant} the spectrum \eqref{penspectrum} reads
\begin{equation}
 H_{\rm lc} = \sum_{i=1}^4 \sum_{n\in \Z}
\sqrt{1+\frac{2\pi^2\lambda}{J^2}n^2}
\hat{N}^i_n+\sum_{a=1}^2\sum_{n\in \Z}\left[
\left(\sqrt{\frac{1}{4}+\frac{2\pi^2\lambda}{J^2}n^2} -
\frac{1}{2}\right) M_n^a +
\left(\sqrt{\frac{1}{4}+\frac{2\pi^2\lambda}{J^2}n^2}+
\frac{1}{2}\right) N_n^a\right] \label{penH}
\end{equation}
We see that this spectrum is consistent with the spectrum found in
\cite{Nishioka:2008gz}. Here we used that from \eqref{dmj} we have that
\begin{equation}
H_{\rm lc} = \Delta - J
\end{equation}

From the spectrum \eqref{penH} we can infer the following dispersion
relation for an $SU(2)\times SU(2)$ magnon
\begin{equation}
\label{disppenrose} \Delta=\sqrt{\frac{1}{4}+\frac{\lambda}{2}p^2}
\end{equation}
where $p=2\pi n/J$ is the momentum of the magnon. This dispersion
relation is clearly consistent with the dispersion relation
\eqref{dispLL} found from the $SU(2)\times SU(2)$ sigma-model limit
as one can see by taking a $p\rightarrow 0$ limit. As explained in
the introduction in Section \ref{sec:intro} this dispersion relation
does not match with weakly coupled ABJM theory.

We can now connect the $SU(2)\times SU(2)$ sigma-model limit of
Section \ref{sec:smlimit} to the above Penrose limit. Consider the
limit $J \rightarrow \infty$. In this limit $c\rightarrow \infty$.
We see therefore from the spectrum \eqref{penH} that the modes
$N^i_n$ and $N^a_n$ decouple, $i.e.$ that the $a^i_n$ and the
$\tilde{a}^a_n$ decouple. Indeed only the $M^a_n$ modes
corresponding to $a^a_n$ are left, giving the spectrum
\begin{equation}
H_{\rm lc} = \sum_{a=1}^2\sum_{n\in \Z} \frac{2\pi^2\lambda}{J^2}n^2
M_n^a \spa \sum_{a=1}^2\sum_{n\in \Z} n M^a_n = 0
\end{equation}
This precisely corresponds to the spectrum of the $SU(2)\times
SU(2)$ sigma-model limit for small $p$, as can be seen from
\eqref{dispLL}. This resembles the $SU(2)$ decoupling limit taken of
the analogous pp-wave solution for the $SU(2)$ sector of type IIB
string theory on $\ads_5\times S^5$
\cite{Harmark:2006ta,Harmark:2008gm}.

We can also connect the above Penrose limit to the $SU(2)\times
SU(2)$ sigma-model limit on the level of the action. Consider the
limit
\begin{equation}
J \rightarrow \infty \spa \frac{\sqrt{J}}{R} x_i = \sqrt{J}
\varphi_i \ \mbox{fixed} \spa \frac{\sqrt{J}}{R} y_i = \sqrt{J}
\theta_i \ \mbox{fixed}
\end{equation}
In this limit we zoom in near a point on each of the two-spheres
that are the target spaces of the double Landau-Lifshitz model
\eqref{LLL}. This gives the action
\begin{equation}
I = \frac{J}{16 \pi^2 \sqrt{2 \lambda} l_s^2 } \sum_{a=1}^2 \int dt'
d\sigma \left[ y_i \partial_{t'} x_i - \frac{\pi^2 \lambda}{J^2} (
{x_i'}^2 + {y_i'}^2 ) \right]
\end{equation}
This is the same action as one obtain by taking a $c \rightarrow
\infty$ limit of the action corresponding to the Lagrangian
\eqref{penlag}.

In conclusion we can connect the $SU(2)\times SU(2)$ sigma-model
limit of Section \ref{sec:smlimit} and the Penrose limit
\eqref{penlim} in the same way as was done in \cite{Harmark:2008gm}
for the $SU(2)$ sector of $\ads_5\times S^5$. In particular, the
above $c\rightarrow \infty$ limit involves a non-relativistic limit
of type IIA string theory on the pp-wave \eqref{ppmet}-\eqref{ppRR}.

\section{New Giant Magnon solution in the $SU(2) \times SU(2)$ sector}
\label{sec:giant}

In this section we find a new Giant Magnon solution in the
$SU(2)\times SU(2)$ sector of type IIA string theory on $\ads_4
\times \C P^3$.

To find the Giant Magnon solution on $AdS_4\times \C P^3$ we
consider the string sigma model on this metric background. The
coordinates can be taken as a 5-vector $Y$ and an 8-vector $X$ where
$X\in S^7$, $Y\in AdS_4$ constrained by
\begin{equation}
\label{s7} X^2= \sum_{i=1}^8X_iX_i=1 \spa Y^2= \sum_{i=1}^3
Y_i^2-Y_4^2 -Y_5^2=-1
\end{equation}
and we furthermore demand
\begin{equation}
\label{cp3s}
C_1 \equiv \sum_{i=1,3,5,7}\left(X_i\partial_\tau X_{i+1}-X_{i+1}\partial_\tau X_i\right)=0\\
 \spa C_2 \equiv \sum_{i=1,3,5,7}\left(X_i\partial_\sigma
X_{i+1}-X_{i+1}\partial_\sigma X_i\right)=0
\end{equation}
defining the background to be $\C P^3$.

 The bosonic part of the sigma model action in the conformal gauge is
\begin{equation}
S= -\sqrt{2\lambda}\int d\tau \int d\sigma \left[
\frac{1}{4}\partial_a Y\cdot \partial^a Y +
\partial_a X\cdot \partial^a X + \tilde \Lambda(Y^2+1)+\Lambda(X^2-1)+\Lambda_1 C^2_1+\Lambda_2 C^2_2\right]
\label{action}\end{equation}
Here, $\Lambda$, $\tilde\Lambda$ and $\Lambda_i$, $i=1,2$ are
Lagrange multipliers which enforce the coordinate constraints
(\ref{s7}) and the constraints (\ref{cp3s}). Keeping into account
the constraints (\ref{cp3s}), the equations of motion following
from the action (\ref{action}) take the form%
\footnote{Here
$\partial^2=\partial_a\partial^a=-\partial_\tau^2+\partial_\sigma^2$.}
\begin{eqnarray}\label{eom}
\left(\partial^2-\Lambda\right)X_i=0~~,~~~~ i=1,\dots, 8\cr
\left(\frac{1}{4}\,\partial^2-\tilde\Lambda\right)Y_i=0~,~~~~i=1,\dots,
5
\end{eqnarray}
and should be supplemented by Virasoro constraints
\begin{eqnarray}\label{vir1}
&&\partial_\tau X\cdot\partial_\tau X+\partial_\sigma
X\cdot\partial_\sigma X+\frac{1}{4}\left(\partial_\tau
Y\cdot\partial_\tau Y+
\partial_\sigma Y\cdot\partial_\sigma Y\right)=0\\ \label{vir2}
&& 2 \partial_\tau X\cdot\partial_\sigma X+\frac{1}{2}\partial_\tau
Y\cdot\partial_\sigma Y=0
\end{eqnarray}
From (\ref{s7}) and (\ref{eom}) it follows that the classical values
of the Lagrange multipliers $\Lambda$ and $\tilde \Lambda$ are
\begin{equation}\label{lagrangemultipliers}
\Lambda=X\cdot\partial^2
X~~,~~~~~~~\tilde\Lambda=-\frac{1}{4}Y\cdot\partial^2 Y
\end{equation}

The Giant Magnon solution will be found as a solution of the
classical equations of motion where only coordinates on two
$S^2\subset S^7$ and $R^1\subset AdS_4$ are excited.  The solution
on $AdS_5\times S^5$ was originally found by Hofman and
Maldacena~\cite{Hofman:2006xt}. This is a closed string solution
with open boundary conditions in one azimuthal direction.

In the case we are studying the solution is point-like in $AdS_4$
and extended along the two $S^2$ which are subsets of $S^7$. The
solution lives on an $R^1\times S^2\times S^2$ subspace of
$AdS_4\times S^7$, the $R^1\subset AdS_4$ and $S^2\times S^2\subset
S^7$. We shall choose the solution in such a way that it has
opposite azimuthal angles in the two $S^2$ and the same polar angle.
The boundary conditions are those of closed string theory. All
variables are periodic, except for the azimuthal angles of the two
$S^2$'s which will be chosen to obey the magnon boundary condition
which on one $S^2$ is
\begin{equation}\label{magnonboundarycondition1}
\Delta \phi_1\equiv p\end{equation} and on the other one will
be\footnote{We denote it $\phi_3$ since the generator of rotations
along the azimuthal direction of the second $S^2$ is called $J_3$.}
\begin{equation}\label{magnonboundarycondition2}
\Delta \phi_3=-p
\end{equation}
These identifications correspond to opposite orientations of the string on the two $S^2$.
 The Giant Magnon is then characterized by the momentum $p$ and
by the choice of the point in the transverse directions to the two $S^2$, $i.e.$ by 2 two-component
polarization vectors. $p$ has to be interpreted as the momentum of the magnon in the spin chain,
these two magnons have equal magnon momentum. They give the same contribution to the total momentum
constraint.

We have found a new solution for the equations (\ref{eom}) satisfying the Virasoro constraints
(\ref{vir1}), (\ref{vir2}) and the constraints (\ref{s7}), (\ref{cp3s}). With our coordinate choice
it reads
\begin{eqnarray}\label{magnonsol}
&&Y_4+iY_5 = e^{i\,2 \, \tau}~~,~~~~Y_1=Y_2=Y_3=0 \\
\label{magnonsolx}&&(X_3,X_4)=\frac{\hat
n_1}{\sqrt{2}}\,\sin\frac{p}{2}\,{\rm sech} u
~~,~~~(X_7,X_8)=\frac{\hat n_2}{\sqrt{2}}\,\sin\frac{p}{2}\,{\rm
sech} u \cr &&X_1+iX_2 = e^{i
\tau}\frac{1}{\sqrt{2}}\left[\cos\frac{p}{2}+i
\sin\frac{p}{2}\,\tanh u\right]\cr&&X_5+iX_6 = e^{-i
\tau}\frac{1}{\sqrt{2}}\left[\cos\frac{p}{2}-i
\sin\frac{p}{2}\,\tanh u\right]
\end{eqnarray}
where $\hat n_i$ $i=1,2$ are two constant unit vectors and
\begin{equation}
u=\Big(\sigma-\tau \cos\frac{p}{2}\Big)\,{\rm csc}\frac{p}{2}
\end{equation}
The Lagrange multipliers (\ref{lagrangemultipliers}) are classically
equal to
\begin{equation}\label{lagmul}
  \Lambda=1-2\,{\rm sech}^2 u ~~,~~~~\tilde\Lambda=1
\end{equation}
as in the usual Giant Magnon solution~\cite{Papathanasiou:2007gd}.

The solution describes right moving solitons traveling along the
worldsheet with velocity $\cos\frac{p}{2}$. The solution on $AdS_4$
in (\ref{magnonsol}) is then chosen so that the energy density,
associated with global time translations, is constant. Rather than
in this energy we are more interested in the conserved quantity
\begin{equation}
\Delta - \frac{J_1-J_3}{2}= -2\sqrt{2\lambda}\int_{-\infty}^\infty
d\sigma\left[ \frac{1}{4}\left(Y_4\dot Y_5-Y_5\dot
Y_4\right)+\frac{X_1\dot X_2-X_2\dot X_1}{2}+\frac{X_6\dot
X_5-X_5\dot X_6}{2}\right]
\end{equation}
where $J_1$ is the charge associated with azimuthal translations on
one of the two $S^2$ and $J_3$ is the generator of the azimuthal
translations on the other $S^2$. The classical value for
$\Delta-J=\Delta - \frac{J_1-J_3}{2}$ on the solution
(\ref{magnonsol})-(\ref{magnonsolx}) then is
\begin{equation}
\label{gmdr} \Delta-J=2\sqrt{2\lambda}\left|\sin\frac{p}{2}\right|
\end{equation}

Note that the above Giant Magnon solution describes two magnons, one for each two-sphere (or
$SU(2)$). Using this fact we can infer from \eqref{gmdr} that the dispersion relation for a single
magnon in the $SU(2)\times SU(2)$ sector of type IIA string theory on $\ads_4 \times \C P^3$ is
\begin{equation}
\label{gmdr2} \Delta-J=\sqrt{2\lambda}\left|\sin\frac{p}{2}\right|
\end{equation}
This dispersion relation is seen to be consistent with the
dispersion relation \eqref{disppenrose} found from the $SU(2)\times
SU(2)$ Penrose limit in Section \ref{sec:penrose}.

\section{Conclusions}
\label{sec:concl}

We studied in this paper the $SU(2)\times SU(2)$ sector in the type
IIA string theory on $\ads_4\times \C P^3$, the proposed string dual
of the recently constructed ABJM theory \cite{Aharony:2008ug}. We
found a sigma-model limit and a Penrose limit corresponding to the
$SU(2)\times SU(2)$ sector and furthermore a new Giant Magnon
solution. Comparing this to the weak coupling results of
\cite{Minahan:2008hf} we found \eqref{dispgen}-\eqref{hl}, showing
that the dispersion relation for ABJM theory has a non-trivial
dependence on $\lambda$.

We note here that beside the dispersion relation
\eqref{dispgen}-\eqref{hl} there are other dispersion relations in
the theory, corresponding to the $\ads_4$ directions and one of the
$\C P^3$ directions. Thus, there might be another independent
interpolation function for these modes.

It would obviously be interesting to study both the dispersion
relations and the S-matrix in the spin chain description for
$\lambda\ll 1$ and $\lambda \gg 1$.

It is also interesting to consider finite-size corrections to the
new Giant Magnon solution found in this paper. This will be
considered in \cite{newgm}.

Finally, we would like to compare with the results of
\cite{Harmark:2008gm}. In \cite{Harmark:2008gm} it was argued that
one can take a $\lambda \rightarrow 0$ limit of type IIB string
theory on $\ads_5 \times S^5$. This limit corresponds to the $SU(2)$
decoupling limit of
\cite{Harmark:2006di,Harmark:2006ta,Harmark:2006ie,Harmark:2007px}.%
\footnote{See also \cite{Astolfi:2008yw}.} It was argued in
\cite{Harmark:2008gm} that in this limit one can quantitatively
match $\CN=4$ SYM with type IIB string theory on $\ads_5 \times
S^5$, and in particular we argued that the one-loop matching was a
result of this. Obviously, this cannot be the case for the duality
between ABJM theory and type IIA string theory on $\ads_4\times \C
P^3$. We believe that the difference between the $\ads_5 /
\mbox{CFT}_4$ case and the $\ads_4 / \mbox{CFT}_3$ is that the
latter duality only possesses 24 supersymmetries. From this we
expect that $\ads_4\times \C P^3$ is not an exact type IIA string
theory background. Indeed, to show that $\ads_5 \times S^5$ is exact
the full 32 supercharges were used \cite{Kallosh:1998qs}. Therefore,
the $\ads_4\times \C P^3$ is indeed more challenging than the
$\ads_5 / \mbox{CFT}_4$ duality.

\section*{Acknowledgments}

We thank Shinji Hirano, Niels Obers and Kostas Zoubos for useful
discussions. TH and MO thank the Carlsberg foundation for support.

\providecommand{\href}[2]{#2}\begingroup\raggedright\endgroup


\end{document}